\documentclass[aps,prb,twocolumn,groupedaddress,showpacs]{revtex4-1}
\usepackage{graphics}
\usepackage{graphicx}
\usepackage{amsmath}
\usepackage{amssymb}
\usepackage{amsthm}
\usepackage{dsfont}
\usepackage{xcolor} 
\begin{document} 
\title        {Some properties of the eigenstates in
the many-electron problem}
\author       {J. Szeftel$^{*}$  and A. Khater$^{+}$}
 \address{
 $^{*}$Laboratoire L\'eon Brillouin(CEA-CNRS), CE-Saclay, 91191 
Gif-sur-Yvette C\'edex, France\\
$^{+}$Laboratoire de Physique des Mat\'eriaux, URA 807, Universit\'e 
du Maine, 72017 Le Mans C\'edex, France}
\begin{abstract}
 A general hamiltonian $H$ of electrons in finite concentration,
 interacting via any  two-body coupling inside 
a crystal of arbitrary dimension, is considered. For simplicity and 
without loss of generality, a one-band model 
is used to account for the electron-crystal interaction. 
The electron motion is described 
in the Hilbert space $S_\phi$, spanned by a basis of Slater 
determinants of one-electron Bloch wave-functions. Electron pairs
 of total momentum $K$ and projected spin $\zeta=0,\pm1$ are   
considered in this work. The hamiltonian then reads 
$H=H_D+\sum_{K,\zeta}H_{K,\zeta}$, where 
$H_D$ consists of the diagonal part of $H$ in the Slater
 determinant basis.  $H_{K,\zeta}$ describes 
the off-diagonal part of the two-electron scattering  process 
which conserves $K$ and $\zeta$.  This hamiltonian
operates in a subspace of $S_\phi$, where the  
Slater determinants consist of pairs 
characterised by the  same $K$ and $\zeta$. 
It is shown that the whole set of eigensolutions $\psi,
\epsilon$ of the 
time-independent Schr\"{o}dinger equation
$(H-\epsilon)\psi=0$ divides in two classes, 
$\psi_1,\epsilon_1$ and $\psi_2,\epsilon_2$. The 
eigensolutions of class 1 are characterised by the property 
that for each solution $\psi_1,\epsilon_1$  there is a single 
$K$ and $\zeta$ such that 
$(H_D+H_{K,\zeta}-\epsilon_1)\psi_{K,\zeta}=0$ where in general 
$\psi_1 \ne \psi_{K,\zeta}$,   
whereas each solution  $\psi_2,\epsilon_2$ of 
class 2 fulfils $(H_D-\epsilon_2)\psi_2=0$.
  We prove also that 
the eigenvectors of class 1 
 have  off-diagonal long-range order  
whereas  those of class 2 do not. Finally our result shows that 
off-diagonal long-range order is not a sufficient condition 
for superconductivity. 
\end{abstract}
 \maketitle
 \begin{center}
\end{center}
                     \section {Introduction}
There has been a long-standing interest for the study of electron 
correlations in  
condensed matter and particularly in the metallic state, because 
these are regarded as playing a paramount role in cooperative 
phenomena such as magnetism and superconductivity. Although
electron correlation is essentially determined
by the Coulomb repulsion effect, three different classes
of models are currently used.\par
	The first is based on the wide efficiency of the
one-electron picture in metals and alloys, fostered by the success of
the Fermi liquid theory \cite{lan}. The electrons behave like a Fermi
gas of independent quasi-particles defined by renormalised parameters 
and finite lifetimes.\par
	The second concerns the magnetic case. It is the realm of 
the repulsive Hubbard model \cite{lon}
and its variations, notably the t-J model
\cite{aue}. As exact results are available only in one dimension
\cite{lie,shib,bares} and for small clusters in two dimensions
\cite{dag}, the groundstate has been approximated by different 
mean-field and  variational procedures,  such as  
Hartree-Fock, Gutzwiller, RVB,
 slave boson state, perturbation \cite{lon,bask,lav} and other 
 calculations. These 
approximations are based on different assumptions, and  
the electron gas is supposed to be either a Fermi liquid of the 
Landau- or Luttinger types\cite{lan,hal}, or a gas giving rise to 
ferromagnetic and antiferromagnetic effects and there is no reliable 
argument 
 to favor either model.\par
	The last concerns the phenomenon of superconductivity. This is 
usually explained within the BCS picture \cite{bar} where
the electrons condense in a variational state 
characterised by off-diagonal long-range order \cite{yan}. 
 The BCS hamiltonian is obtained by  
truncating an attractive Hubbard hamiltonian
in reciprocal space. Consequently  the BCS hamiltonian, once
Fourier-transformed back to real space, turns out to display four
site, interelectron coupling terms which are not present in 
the Hubbard hamiltonian, 
used to describe electron interactions in the normal state.\par
	Although the three  above classes employ  different 
hamiltonians, the Hilbert space is in all cases taken to be based on  
Slater determinants and is designated here as $S_\phi$.\par
	Our work
investigates the properties of the eigenstates of a general  
many-body hamiltonian $H$. We present a  mathematical proof 
that the set of eigenstates of $H$ in $S_\phi$, including  
in particular  the groundstate, divides in two classes 
 $\psi_1$ and $\psi_2$ which differ 
by their off-diagonal long-range order properties.
 These results are
valid for any electron concentration and arbitrary crystal dimension, 
and for any  
interelectron coupling provided it is of a  two-body nature. The proof 
exploits specifically the property of the conservation of the pair 
momentum in every two-electron scattering
event. An approximation, consisting of dealing with such pairs as if 
they 
were  independent quasi-particles \cite{sz1}, has already provided
the groundstate energy of the one-dimensional Hubbard model in
excellent agreement with the exact result \cite{shib}.
 In the general case of arbitrary dimension and general hamiltonian 
 investigated in 
this work,  it is necessary to introduce an auxiliary
 Hilbert space $S_{\otimes\phi}$ in order to derive the $\psi_1$ or 
$\psi_2$ like properties of the eigenstates. 
 $S_{\otimes\phi}$  is built over a set of pairs characterised
by their total momentum $K$ and projected spin $\zeta$. 
Other authors \cite{bar,gir,bare} 
have also used
such sets, nevertheless they remained within the framework of
$S_\phi$.\par 
	The outline is as follows: in section 2 the many-body
hamiltonian $H$ is presented and the problem to be solved is
set out; section 3 provides the definition of the auxiliary 
space $S_{\otimes\phi}$ as
well as its algebraic properties; sections 4 and 5 detail the proofs 
of  two
Theorems establishing the either $\psi_1$ like and $\psi_2$ like 
properties 
of the eigenstates of
$H$ in the usual space of Slater determinants $S_\phi$ 
(a partial account of section 4 has been published
elsewhere \cite{sz2}); the physical consequences of these
results are summarised in the concluding section 6.
	\section {The many-body hamiltonian}
	In the following model we consider a crystal  
containing $N$ sites and $2n$ itinerant electrons where $N>>1$ 
and $n>>1$. 
The crystal can have  arbitrary dimension. These electrons populate 
a single band where the one-electron energy reads $E(k)$ and $k$ is a
vector of the Brillouin zone. To simplify the discussion  and 
without loss of generality, we consider that $E(k)$ is independent of
the electron spin $\sigma=\pm1/2$. The Pauli principle requires that 
$n\leq N$. Let the electrons be coupled via a  spin 
independent pair potential $V$.
 The total system hamiltonian $H$ can be written in reciprocal space
  as:  
\begin{equation} 
\label{h0}
H=\sum_{k,\sigma}E(k)c^+_{k,\sigma}c_{k,\sigma}\hspace{1ex}+
\sum_{K,k,k',\sigma_{i=1,..4}}V(K,k,k')
c^+_{k,\sigma_1}c^+_{K-k,\sigma_2}c_{K-k',\sigma_3}c_{k',
\sigma_4}\hspace{1ex},  
\end{equation} 
where the first term denotes the one-electron contribution  and the 
second 
denotes the most general expression to describe two-body interactions 
in 
a periodic crystal. 
The operators $c^+_{k,\sigma}$  and $c_{k,\sigma}$ 
are  one-electron creation and  annihilation  operators on the Bloch 
state
$k,\sigma$. They obey the usual Fermi  commutation rules.  
The real coefficients $V(K,k,k')$ are the matrix elements of the  
two-electron scattering process, conserving 
the momentum $K$ of each scattered pair.  For usual pair potentials 
involving only two-site terms in real space, $V(K,k,k')$ is  
 $K$-independent and depends only of $(k-k')$. The 
summations in eq.\ref{h0} are carried out over all possible values of  
$K,k,k'$ in the Brillouin zone 
under the constraint of spin conservation 
 $\sigma_1+\sigma_2=\sigma_3+\sigma_4$ $(\sigma_{i=1,..4}=\pm1/2)$.
 A special case of eq.\ref{h0} is the Hubbard hamiltonian \cite {h1} 
 which 
is  recovered by setting
$E(k)=\cos(\sum k_i)$ where the components of $k$ are identified 
by $k_i$, 
 $\sigma_1+\sigma_2=0$ and 
$V(K,k,k')$ is  a constant $U/N$ for all scattering events. 
The hamiltonian $H$ describes the electron  motion 
in the Hilbert space $S_\phi$ of dimension $d_{\phi}=
\left(\begin{array}{c}2N\\2n\end{array}\right)$. Each basis vector 
 $\phi_i$ with $i=1,..d_{\phi}$ is a Slater determinant
involving $2n$ one-electron Bloch states.\par
	Since this discussion resorts repeatedly to electron pairs, 
it is convenient to introduce the following pair creation and 
annihilation
operators $b^+_{\zeta}(k,k'),b_{\zeta}(k,k')$:  
\begin{equation}
\label{pair}
\begin{array}{c}
b^+_{\pm1}(k,k')=c^+_{k,\pm}c^+_{k',\pm}\hspace{1em},
\hspace{1em}b_{\pm1}(k,k')=c_{k',\pm}c_{k,\pm}
\hspace{1em},\\\hspace{1em}
b^+_0(k,k')=c^+_{k,+}c^+_{k',-}\hspace{1em},
\hspace{1em}b_0(k,k')=c_{k',-}c_{k,+}\hspace{1em}.
\end{array}
\end{equation} 
The subscripts $+$ or $-$ in the one-electron $c^{(+)}_{k,\pm}$ operators 
refer to the two possible directions of the electron spin. 
The subscript $\zeta=0,\pm1$ stands for the projection of the total 
spin 
of the pair where $\zeta=\pm1$ indicates the same spin   
and $\zeta=0$ indicates opposite spins on both electrons, 
before and after scattering. 
The commutation rules of such pairs are  neither  
Fermi- nor Bose-like. It is useful to recast the hamiltonian $H$ of  
 eq.\ref{h0} in terms 
 of  the subsidiary hamiltonians $H_D$,  $H_{K,\zeta}$ as follows:
\begin{equation} 
\label{h2}
H= H_D+\sum_{K,\zeta=0,\pm 1}H_{K,\zeta}\hspace{2ex},
\end{equation}
where $H_D$ and  $H_{K,\zeta}$ may be written as:
\begin{equation} 
\label{h1}
\begin{array}{c}
H_D=\sum_{k,\sigma}E(k)c^+_{k,\sigma}c_{k,\sigma}
+\sum_{k,k'}V(k+k',k,k)c^+_{k,+}c_{k,+}
c^+_{k',-}c_{k',-}\\
+\sum_{k,k',\sigma}(V(k+k',k,k)-V(k+k',k,k'))
c^+_{k,\sigma}c_{k,\sigma}
c^+_{k',\sigma}c_{k',\sigma}\hspace{2ex},\\ H_{K,0}=\sum_{k,k'\ne
k}V(K,k,k')b^{+}_{0}(k,K-k)b_{0}(k',K-k')\hspace{2ex},\\
H_{K,\pm 1}=\sum_{k,k'\ne (k,K-k)}V(K,k,k')b^{+}_{\pm
1}(k,K-k)b_{\pm 1}(k',K-k')\hspace{2ex}.  
\end{array} 
\end{equation}
The diagonal matrix elements of $H$ in the Slater determinant
 basis are regrouped in the hamiltonian $H_D$. 
 Inversely the off-diagonal matrix
elements of $H$ are regrouped in the hamiltonians 
$H_{K,\zeta}$.  In the Hubbard hamiltonian, $H_D$ 
takes the form $\sum_{k,\sigma}E(k)c^+_{k,\sigma}c_{k,\sigma}
+\frac{U}{N}\sum_{k,k'}c^+_{k,+}c_{k,+}
c^+_{k',-}c_{k',-}$ and $h_{K,\pm1}=0$ 
for every $K$. Note also that the BCS hamiltonian 
reads as $H_D+H_{K=0,\zeta=0}$ where $H_D$ and 
$H_{K=0,\zeta=0}$ are given by their particular expressions in 
the Hubbard hamiltonian.\par
	The main purpose of this article is to present and demonstrate
 two Theorems which 
 characterise the two classes of eigensolutions $\psi,\epsilon$ of the 
time-independent Schr\"{o}dinger equation
$(H-\epsilon)\psi=0$ where $H$ is given by eq.\ref{h0} and $\psi$  
belongs to the Hibert space $S_\phi$. These classes are designated 
respectively as $\psi_1,\epsilon_1$ and $\psi_2,\epsilon_2$.\\
 {\bf Theorem 1 : \it To each eigensolution $\psi_{K,\zeta},
 \epsilon_1$ 
where $(H_D+H_{K,\zeta}-\epsilon_1)\psi_{K,\zeta}=0$, 
there corresponds an eigensolution $\psi_1,\epsilon_1$ of $H$ such 
that 
$(H-\epsilon_1)\psi_1=0$.}\\
The fingerprint of each $\psi_1$ is that its linear expansion 
 over the basis vectors of 
$S_\phi$ involves at least one Slater 
determinant $\phi$ which 
can be written as:
\begin{equation} 
\label{nK} 
\phi=\prod_{j=1}^n
b^+_\zeta(k_j,K-k_j)|0\rangle\hspace{2ex}, 
\end{equation}
where $|0\rangle$ designates the no-electron state.  
Note that $\psi_{K,\zeta}$ in general is not an eigenvector of $H$ 
although $\epsilon_1$
 is indeed an eigenvalue of $H$.\\
{\bf Theorem 2 : \it For every $\psi_2,\epsilon_2$, the equation 
$(H-\epsilon_2)\psi_2=0$
     implies that  
$(H_D-\epsilon_2)\psi_2=0$.}\\
Each $\psi_2$ is characterised by its linear 
expansion over the basis vectors of 
$S_\phi$ containing no Slater determinant such as $\phi$ in 
eq.\ref{nK} for every $K$ and $\zeta$.\par
	In the simple case of a two-electron system, 
that is  a single pair 
$(n=1)$,
  Theorem 1 has been demonstrated previously \cite{h1}. This  result 
  follows since 
 $H$ and $H_{K,\zeta}$ commute with each other and with 
the pair number operator  $N_{K,\zeta}$:  
\begin{equation}
\label{N} 
N_{K,\zeta}=\sum_k
b^+_\zeta(k,K-k)b_\zeta(k,K-k)\hspace{1em}.  
\end{equation}
Our aim hence is  to generalise the result of reference \cite{h1} 
to the $n>1$ case.
 While it is easy to show that $H_{K,\zeta}$ and $N_{K,\zeta}$
 still commute
for any $n$, the operators $H$ and $N_{K,\zeta}$ however no longer
commute in this general case. Therefore the $n>1$ case cannot be 
dealt with in the  
Hilbert space $S_\phi$ of Slater determinants. It becomes then 
necessary to 
treat the problem in  
an auxiliary Hilbert space $S_{\otimes\phi}$ which is purposely 
constructed so that $H$ and $N_{K,\zeta}$ commute in this space, 
keeping invariant their definitions as in eq.\ref{h0} and 
eq.\ref{N}.\par
	As $\psi_1$ eigenstates will be shown 
in addition to have off-diagonal long range order whereas 
 $\psi_2$ eigenstates do not have, it is in order to recall the 
definition of the two-body correlation function attached to 
this particular kind of long range order 
characterising the BCS state: 
\begin{equation} 
\label{OD} 
f_{odlro}(|\tau|)=\sum_{i,j,l,m,\sigma_{h=1,..4}}\langle
\phi|c^+_{i,\sigma_1}c^+_{j,\sigma_2}c_{l,\sigma_3}
c_{m,\sigma_4}|\phi\rangle\hspace{1ex}, 
\end{equation}
where the Wannier operator $c^{(+)}_{i,\sigma_h}$ destroys (creates) an
electron with spin $\sigma_h$ at site $i$ labeled by the lattice vector
$r_i$ and the sum is done with 
$(r_j-r_i)=(r_m-r_l)=\rho$, $(r_i-r_l)=\tau$ and   
$\sigma_1+\sigma_2=\sigma_3+\sigma_4$. 
Eq.\ref{OD} extends to the $\rho\ne0$ case the  
usual definition of off-diagonal long-range order \cite{yan} given 
 in the Hubbard model for $\rho=0$ and $\sigma_1=-\sigma_2$. 
 A many-electron  
state $\phi \in S_\phi$ is said to have off-diagonal long range order 
if $f_{odlro}(|\tau|)$, calculated 
at $\rho$ kept fixed, oscillates versus $|\tau|$ without decaying 
to zero for $|\tau|\rightarrow\infty$. It must be noticed that 
 off-diagonal long range order differs from real space long 
range order, typical of crystalline matter, magnetic materials, spin-
 and charge-density waves. This latter type of long range order 
is characterised by the following two-body correlation function:
\begin{equation} 
\label{RS} 
f_{rslro}(|\tau|)=\sum_{i,j,\sigma_{h=1,..4}}\left(
\langle\phi|c^+_{i,\sigma_1}c_{i,\sigma_2}c^+_{j,\sigma_3}
c_{j,\sigma_4}|\phi\rangle-\langle\phi|c^+_{i,\sigma_1}
c_{i,\sigma_2}|\phi\rangle\langle\phi|c^+_{j,\sigma_3}
c_{j,\sigma_4}|\phi\rangle\right)\hspace{1ex}, 
\end{equation}
where the sum is done with $(r_i-r_j)=\tau$ and 
$\sigma_1+\sigma_3=\sigma_2+\sigma_4$. Charge and spin 
fluctuations correspond  respectively to $\sigma_1=\sigma_2$ 
 and $\sigma_1=-\sigma_2$. 
A state $\phi \in S_\phi$ is said 
to have real space long range order 
if $f_{rslro}(|\tau|)$ oscillates versus $|\tau|$ without decaying 
to zero for $|\tau|\rightarrow\infty$. By comparing the 
definition in eq.\ref{OD} with that in 
eq.\ref{RS}, it is realized that $f_{odlro}(|\tau|)\ne f_{rslro}
(|\tau|)$ even if $\rho=0$. Besides from the experimental point 
of view, real space long range order gives rise to Bragg difffraction 
in a neutron or X-ray scattering experiment while off-diagonal 
long range order  does not.
         \section{Properties of the auxiliary Hilbert space 
         $S_{\otimes\phi}$ }
Any Slater determinant $\phi_e$ of $S_\phi$ can be written as:
\begin{equation}
\label{Sla}
\phi_e=\prod_{K,\zeta}\left(\prod_{j=1}^{n_{K,\zeta}}
b^+_{\zeta}(k_j,K-k_j)\right)|0\rangle\hspace{2ex},
\end{equation} 
where  all  pairs  
$b^+_\zeta(k_j,K-k_j)|0\rangle$ having the same $K$  and 
$ \zeta$ have been regrouped together. In the product with 
respect to  
the index $j$, the $e$ dependence of $j$ has been dropped 
for simplicity. The integer 
$n_{K,\zeta}\ge 0$ designates the 
total number of pairs characterised by $K, \zeta$ in $\phi_e$, and 
the $n_{K,\zeta}$'s 
satisfy $\sum _{K,\zeta} n_{K,\zeta}=n$. The basis vector 
$\Phi_{e,\alpha}$ 
of $S_{\otimes\phi}$ is defined from $\phi_e$ as:
\begin{equation} 
\label{30}
 \Phi_{e,\alpha}=\bigotimes_{K,\zeta} \phi_{K,\zeta}\hspace{2eM},
\hspace{2eM}\phi_{K,\zeta}=\prod_{j
=1}^{n_{K,\zeta}}
b^+_{\zeta}(k_j,K-k_j)|0\rangle\hspace{2eM},  
\end{equation} 
where the tensor product replaces the simple  product 
$\prod_{K,\zeta}$
of eq.\ref{Sla} and each $\phi_{K,\zeta}$ is a Slater determinant
containing  $n_{K,\zeta}$ of pairs $K, \zeta$. The sequence of 
integers 
$\{n_{K,\zeta}\}$ in eqs.\ref{Sla},\ref{30} defines uniquely the 
pair 
configuration $\alpha$ of $\phi_e$. Therefore $n_{K,\zeta}$ will be 
denoted 
$n_{K,\zeta,\alpha}$ in the following.
 The set of  pair configurations of $\phi_e$ can be obtained 
by selecting $m$ permutations of $2n$  one-electron 
Bloch states defining $\phi_e$. The number of pair 
configurations $m=(2n)!/(2^{n}(n!))$ 
is smaller than that of permutations $(2n)!$ because many 
different permutations correspond to the same pair configuration.
 The basis vectors $\Phi_{e,\alpha}$ of $S_{\otimes\phi}$ are 
 generated by 
letting the subscripts $e=1,..d_\phi$ and $\alpha=1,..m$ run over 
all possible
 values, which implies that the dimension of $S_{\otimes\phi}$ is 
 equal to 
 $md_\phi$. 
 The pair number operator 
$N_{K,\zeta}$  
is taken to act on $\Phi_{e,\alpha}$ as follows:
\begin{equation} 
\label{N1}
N_{K,\zeta}\Phi_{e,\alpha}=
n_{K,\zeta,\alpha}\Phi_{e,\alpha}\hspace{1eM}.
\end{equation} 
 As the $\Phi_{e,\alpha}$'s are chosen to be orthonormal,
eq.\ref{N1} entails that $n_{K,\zeta,\alpha}=
\langle\Phi_{e,\alpha}|N_{K,\zeta}|\Phi_{e,\alpha}\rangle$.\par
	The subspace $S_\Phi \subset S_{\otimes\phi}$  is then
 introduced as spanned by the basis vectors $\Phi_e$ defined by:
\begin{equation} 
\label{exp}
\Phi_e=\sum_{\alpha=1}^{m}\Phi_{e,\alpha}
\hspace{2ex},
\end{equation} 
where the sum is carried over $m$ pair configurations $\alpha$ of 
$\phi_e$. 
The one to one correspondence between $\phi_e \in S_\phi$ 
and $\Phi_e \in S_\Phi$ ensures that the dimension of 
$S_\Phi$ is equal to $d_\phi$. Although  
$S_\Phi$ and $S_{\otimes\phi}$ obey the Pauli principle by 
construction, 
the vectors $\Phi_e \in S_\Phi$ and $\Phi_{e,\alpha} \in 
S_{\otimes\phi}$ 
do not exhibit 
the antisymmetry property typical of Slater determinants with respect 
to interchanging two electrons. The question of redundancy,  
encountered here, since the dimension of
$S_{\otimes\phi}$ is larger than that of $S_\Phi$, arises as in other 
works \cite{gir,bare} dealing with
electron pairs. However in our treatment this redundancy does  
not pose any particular problem.
 The significance of $\Phi_e,\Phi_{e,\alpha},
n_{K,\zeta,\alpha}$ is illustrated in detail in the appendix  for 
the exemplifying case of a four electron system.\par
	Introduce now the subspaces 
 $S_{K,\zeta}\subset S_\Phi$ and $S_2\subset S_\Phi$, where 
 $S_{K,\zeta}$  is defined for each $K,\zeta$ as spanned
 by the basis vectors 
$\Phi_{i=1,..d_\zeta}$, $d_\zeta$ being the dimension 
of $S_{K,\zeta}$. By definition each  $\Phi_i$ is associated 
with a Slater determinant $\phi_i$ of $S_\phi$, such as 
in eq.\ref{nK} and thus comprising  
$n$ pairs, all having the same $K$ and 
 $\zeta$. The dimension $d_\zeta$ of $S_{K,\zeta}$
is $d_0=\left(\begin{array}{c}N\\n\end{array}\right)$ or 
$d_{\pm1}=\left(\begin{array}{c}\frac{N}{2}\\n\end{array}\right)$ 
depending whether $\zeta=0$ or $\zeta=\pm1$, respectively. 
The  characteristic property of each $\Phi_i$ is that its pair 
configuration expansion, as  given in 
eq.\ref{exp}, involves a particular value $\gamma$ defined by:
\begin{equation} 
\label{alpha}
\Phi_i=\sum_{\alpha=1}^{m}\Phi_{i,\alpha}\hspace{1ex},\hspace{1ex}
n_{K,\zeta,\gamma}=\langle\Phi_{i,\gamma}
|N_{K,\zeta }|\Phi_{i,\gamma}\rangle=n \Rightarrow
n_{K',\zeta',\gamma}=\langle\Phi _{i,\gamma}|N_{K',\zeta' }|
\Phi_{i,\gamma}\rangle=0\hspace{1ex}, 
\end{equation}
where $K'$ and $\zeta'$ take all possible values different from $K$ 
and $\zeta$ respectively. Each 
$\Phi_{i,\gamma}$ of $S_{\otimes\phi}$ is then written in  
the same expression as $\phi_{i}\in S_\phi$
 of eq.\ref{nK}, associated  with $\Phi_{i}\in S_{K,\zeta}$, 
because the tensor product yielding 
$\Phi_{i,\gamma}$ as in eq.\ref{30} reduces to a single 
Slater determinant of $n$ pairs $K,\zeta$. 
Inversely the subspace $S_2$  is 
spanned by the basis vectors $\Phi_{p=1,..d_2}$ of $S_\Phi$,
 $d_2$ being the dimension of $S_2$ ($S_2$ is named so because it 
will be shown hereafter to include all  $\psi_2$ like eigenvectors). 
Each $\Phi_p$ is characterised by:
\begin{equation} 
\label{S2}
\Phi_p=\sum_{\beta=1}^{m}\Phi_{p,\beta}
\hspace{1ex},\hspace{1ex}
n_{K,\zeta,\beta}=\langle\Phi_{p,\beta}
|N_{K,\zeta }|\Phi_{p,\beta}\rangle<n\hspace{1ex},\hspace{1ex}
\forall \hspace{1ex} K,\zeta \hspace{1ex},
 \end{equation}
where the inequality holds for every $\beta$ value involved in
 the pair configuration expansion 
of $\Phi_p$. As the 
subspaces $S_2$ 
and $S_{K,\zeta}$ are disjoint, because their characteristic 
properties 
as expressed by eqs.\ref{alpha},\ref{S2} exclude one another,
 they provide a basis for $S_\Phi$:
\begin{equation} 
\label{SD}
 S_\Phi=S_2\bigoplus_{K,\zeta}S_{K,\zeta}\hspace{1ex},\hspace{1ex}
d_\phi=d_2+N(d_0+2d_{\pm1})\hspace{1ex}.
\end{equation}
\par	Consider now the following expression for the 
hamiltonian $H'$ in $S_{\otimes\phi}$: 
\begin{equation} 
\label{HSr}
H'=\sum_{i,j}\langle \phi_i|H|\phi_j\rangle|
\Phi_{i,\gamma}\rangle\langle\Phi_{j,\gamma}| +
\sum_{p,q,\beta}m_{pq}\langle \phi_p|H|\phi_q\rangle|
\Phi_{p,\beta}\rangle\langle\Phi_{q,\beta}|\hspace{2ex},
\end{equation} 
where the sum with respect to $i,j$ is performed on all 
Slater determinants  $ \phi_i $ and $\phi_j$ associated respectively 
with 
$ \Phi_i \in S_{K,\zeta}$ and $\Phi_j\in S_{K,\zeta}$, the pair 
configuration $\gamma$ is defined in eq.\ref{alpha} and $K,\zeta$ 
take all possible values. The sum with 
respect 
to $p,q$ is carried over all  $ \Phi_p$ and 
$ \Phi_q$ such that $ \Phi_p$ or $ \Phi_q$ belong to 
$S_2$.  The matrix elements $\langle \phi_i|H|\phi_j\rangle$ and 
$\langle \phi_p|H|\phi_q\rangle$ are calculated with $H$ given by 
eq.\ref{h0}.
 As $\langle \phi_p|H|\phi_q\rangle\ne 0$ requires that the 
Slater determinants  $\phi_p$ 
and $\phi_q$ differ by one pair only, and  they  read  
$\phi_p=b^{+}_\zeta(k,K-k)\phi_{pq}|0\rangle$ and
$\phi_q=b^{+}_\zeta(k',K-k')\phi_{pq}|0\rangle$ where 
 $\phi_{pq}$ comprises the product of $(n-1)$ pairs, 
the sum with respect 
to $\beta$ is made with $m_{pp}=1/m$ and 
$m_{pq}=(2n-1)/m$ over $m/(2n-1)$ pair configurations 
 common to $ \Phi_p $ and $\Phi_q$. The definition of $H'$ 
ensures that  
the matrix elements $\langle \Phi_e|
H'|\Phi_f\rangle$ and 
$\langle \phi_e|H|\phi_f\rangle$ are equal 
 for all $e,f$ values 
where $ \phi_e ,\phi_f$ are two Slater determinants of $S_\phi$  and 
$ \Phi_e ,\Phi_f$ are the corresponding basis vectors of $S_\Phi$.  
It follows that the Schr\"{o}dinger equations 
$(H-\epsilon)\psi=0$ and $(H'-\epsilon)\Psi=0$, where 
$\psi\in S_\phi$ and $\Psi\in S_\Phi$, have the same 
spectrum of 
eigenvalues $\epsilon$ and there is a  one to one 
correspondence between $\psi$ and $\Psi$.\par 
	 Since $H'$ in eq.\ref{HSr} does not display such terms as 
$|\Phi_{p,\alpha}\rangle\langle\Phi_{q,\beta}|$ which would 
mix two different pair configurations $\alpha$ and $\beta$, the 
Schr\"{o}dinger equation $(H'-\epsilon)\Psi=0$, where $\Psi$ 
belongs to $S_\Phi$,
splits into partial Schr\"{o}dinger equations:
\begin{equation} 
\label{split}
(H'-\epsilon)\Psi=0\hspace{1ex}, \hspace{1ex}
\Psi=\sum_{e =1}^{d_\phi}a_e\Phi_e\hspace{1ex}, \hspace{1ex}
\Phi_e=\sum_{\alpha=1}^{m}\Phi_{e,\alpha}\Rightarrow
(H'-\epsilon)\Psi_\alpha=0\hspace{1ex}, \hspace{1ex}
 \Psi_\alpha=\sum_{e=1}^{d_\phi}a_e\Phi_{e,\alpha}
\hspace{1ex}, \hspace{1ex}
\Psi=\sum_\alpha \Psi_\alpha\hspace{1ex},
\end{equation}
where the coefficients $a_e$ are real, the sum over  
$\alpha$ is the pair configuration expansion of $\Phi_e$ and 
 $\Psi_\alpha$ belongs to $S_{\otimes\phi}$.
	      \section {Proof of Theorem 1}
Consider the Schr\"{o}dinger equation 
$(H'-\epsilon_1)\Psi_1=0$ where 
the eigenvector $\Psi_1 \in S_\Phi$  
is assumed to have a non vanishing projection in $S_{K,\zeta}$ and   
thus  reads:
\begin{equation} 
\label{T1}
\Psi_1=\Psi_{K,\zeta}+\Psi_1'\hspace{1ex},\hspace{1ex}
\Psi_{K,\zeta}=\sum_{i=1}^{d_\zeta}a_i\Phi_i
\hspace{1ex},\hspace{1ex}
\Psi_1'=\sum_{p=1}^{d_2}a_p\Phi_p\hspace{1ex},
\end{equation} 
where the coefficients $a_i,a_p$ are real and the $\Phi_i$'s and 
$\Phi_p$'s are basis vectors of $S_{K,\zeta}$ and $S_2$, respectively. 
We now apply eq.\ref{split} to $\Psi_1$ for the particular pair 
configuration $\gamma$ defined in eq.\ref{alpha}: 
\begin{equation} 
\label{T2}
(H'-\epsilon)\Psi_{1,\gamma}=0\hspace{1ex}, \hspace{1ex}
 \Psi_{1,\gamma}=\Psi_{K,\zeta,\gamma}+\Psi'_{1,\gamma}
\hspace{1ex}.
\end{equation} 
As the vector $\Psi'_1$ 
is inferred from the definition of $\Phi_p$ in eq.\ref{S2} 
not to contribute to 
$\Psi_{1,\gamma}$, 
it ensues that $\Psi_{1,\gamma}$ reduces to $\Psi_{K,\zeta,\gamma}$.   
Because of  $\langle\phi_i|H|\phi_j\rangle
=\langle\phi_i|H_{D}+H_{K,\zeta }|\phi_j\rangle$ which holds for the
 hamiltonians $H_{D}$ and $H_{K,\zeta }$ in eq.\ref{h1}  and any two  
Slater determinants $\phi_i,\phi_j$ associated  with the basis vectors
 $\Phi_i,
\Phi_j$ of $S_{K,\zeta}$, it comes finally:
\begin{equation} 
\label{psi1}
\left(H'-\epsilon_1\right)\Psi_{1,\gamma}=0\Rightarrow 
\left(H_{D}+H_{K,\zeta }-\epsilon_1\right)
\Psi_{K,\zeta,\gamma}=0\Leftrightarrow
\left(H_{D}+H_{K,\zeta }-\epsilon_1\right)
\psi_{K,\zeta}=0\hspace{1ex},
\end{equation}
where $\psi_{K,\zeta}\in S_\phi$ is in one to one correspondence 
with $\Psi_{K,\zeta}\in S_\Phi$. Eq.\ref{psi1} means that, if  
$(\psi_{K,\zeta}+\psi'_1)$ and $\epsilon_1$ are eigenvector and 
eigenvalue 
of $H$ in $S_\phi$, the vector $\psi_{K,\zeta}$ and $\epsilon_1$ 
are eigenvector 
and eigenvalue 
of $\left(H_D+H_{K,\zeta }\right)$ in $S_\phi$ too.
 To complete the proof of Theorem 1 it must be shown in addition 
that every eigensolution 
 $\psi_{K,\zeta},\epsilon_1$ of $\left(H_D+H_{K,\zeta }\right)$ 
gives rise to an eigensolution $\psi_1,\epsilon_1$ of $H$. 
The latter will be  proved now 
by contradiction. Suppose that there is an eigenvalue of some 
hamiltonian 
$\left(H_D+H_{K,\zeta}\right)$ which is not an eigenvalue of $H$. 
Then the corresponding $S_{K,\zeta }$ will contribute only 
$(d_\zeta-1)$ 
eigenvalues instead
 of $d_\zeta$ to the spectrum of $H$, which will result in an 
 uncomplete 
diagonal basis for $H$ and is thus at odds with 
the property of  $H$ being hermitian. {\bf Q.E.D.}\par
	 Both  $\psi_{K,\zeta }$ and the BCS variational 
state \cite{bar}
consist of a linear combination of Slater determinants of pairs having
the same $K,\zeta$. They differ, however, by the number of pairs in 
each
determinant, which ranges from $0$ up to $N$ in the BCS state
while it is always equal to $n$ for $\psi_{K,\zeta }$.
 As for the BCS state \cite{yan},  
 off-diagonal long-range order, as defined in eq.\ref{OD}, 
is a fingerprint of $\psi_{K,\zeta}$ :
\begin{equation} 
\label{BCS} 
f_{odlro}(|\tau|)=\cos(K.\tau)\Delta\hspace{1ex},
\hspace{1ex}\Delta=\sum_{k,k'}e^{i((k-k').\rho)}\langle
\psi_{K,\zeta }|b^+_\zeta(k,K-k)b_\zeta(k',K-k')|
\psi_{K,\zeta }\rangle\hspace{1ex}, 
\end{equation}
where $\Delta$ 
is a two-body correlation parameter attached to $\psi_{K,\zeta }$. 
 Actually $f_{odlro}(|\tau|)$ results
from a sum over all $K',\zeta'$ but the contributions with
$(K',\zeta')\ne (K,\zeta)$ vanish identically. 
$f_{odlro}(|\tau|)$ oscillates without 
decaying for $|\tau|\rightarrow\infty$ provided $\Delta\ne0$.  It will  
be shown in the following section that $\psi'_1$ (see eq.\ref{T1}) 
contributes nothing to $f_{odlro}(|\tau|)$ for 
$|\tau|\rightarrow\infty$ 
so that $\psi_{K,\zeta }$ and $\psi_1$ have the 
same off-diagonal long-range order parameter. 
In the Hubbard model, the validity of Theorem 1 has 
been confirmed \cite{yan}
 for a large class of many-electron eigenstates $(K=(\pi,\pi,\pi)
,\zeta=0)$, built with help of 
the $\eta$-
pairing mechanism,  for arbitrary interelectron coupling $U$ and 
electron concentration.
	     \section {Proof of Theorem 2}
	 We turn now to the Schr\"{o}dinger equation  
$(H'-\epsilon_2)\Psi_2=0$ where $H'$ is given by eq.\ref{HSr} and 
the eigenvector $\Psi_2$ 
belongs to the subspace $S_2$ of $ S_\Phi$:
\begin{equation} 
\label{t1}
\Psi_2=\sum_{p=1}^{d_2}a_p\Phi_p\hspace{1ex},\hspace{1ex}
\Phi_p=\sum_{\beta=1}^{m}\Phi_{p,\beta}
\hspace{1ex},\hspace{1ex}
\Psi_{2,\beta}=\sum_{p=1}^{d_2}a_p\Phi_{p,\beta}
\hspace{1ex},\hspace{1ex}
\Psi_2=\sum_\beta\Psi_{2,\beta}\hspace{1ex},
\end{equation} 
where the $a_p$'s are real and the pair configuration expansion of  
$\Phi_p$ is done with respect to $\beta$.
To demonstrate the validity of Theorem 2 it is sufficient to show that 
the matrix 
element $\langle\Phi_p|H'|\Phi_q\rangle$ vanishes for  
all $\Phi_p$ and $\Phi_q$ in the linear expansion giving $\Psi_2$ 
in eq.\ref{t1} if $p\ne q$.  The 
proof proceeds by contradiction. Suppose that 
$\langle\Phi_p|H'|\Phi_q\rangle\ne0$ for $p=1$ and $q=2$ 
whereas $\langle\Phi_p|H'|\Phi_q\rangle=0$ for $p\ne1,2$, 
$q\ne 1,2$ and $p\ne q$. This implies for the  Schr\"{o}dinger   
equation $(H'-\epsilon_2)\Psi_2=0$:
\begin{equation} 
\label{t2}
(H'-\epsilon_2)(a_1\Phi_1+a_2\Phi_2)+
\sum_{q\ne1,2}(\langle\Phi_q|H_D|\Phi_q\rangle-\epsilon_2)
a_q\Phi_q=0\hspace{1ex}.
\end{equation} 
As the basis vectors $\Phi_q$ are linearly independent, eq.\ref{t2} 
implies that:  
\begin{equation} 
\label{t3}
(H'-\epsilon_2)(a_1\Phi_1+a_2\Phi_2)=0\hspace{1ex},\hspace{1ex}
\langle\Phi_q|H_D|\Phi_q\rangle=\epsilon_2\hspace{1ex},
\forall q\ne 1,2\hspace{1ex}.
\end{equation} 
As seen in eq.\ref{HSr},  
$\langle\Phi_1|H'|\Phi_2\rangle\ne0$ requires that $\Phi_1$ 
and $\Phi_2$ differ by one pair only so that  they  read  
$\Phi_1=b^{+}(k_1,k_2)|0\rangle\otimes\Phi_{12}$ and
$\Phi_2=b^{+}(k_3,k_4)|0\rangle\otimes\Phi_{12}$ where 
the spin index $\zeta$ is
dropped for simplicity till the end of this proof, 
$k_1+k_2=k_3+k_4$ and $\Phi_{12}$ includes 
$(n-1)$ pairs. Moreover due to eq.\ref{split}, 
the expression $(H'-\epsilon_2)(a_1\Phi_1+a_2\Phi_2)=0$ 
in eq.\ref{t3} splits in $S_{\otimes\phi}$ into partial 
Schr\"{o}dinger equations 
$(H'-\epsilon_2)(a_1\Phi_{1,\beta}+a_2\Phi_{2,\beta})=0$ where 
the pair configuration index $\beta$ runs over all values allowed 
by eq.\ref {t1}. The particular case of 
$\beta$, where the pair numbers $n_{k_1+k_2,\beta}=
n_{k_3+k_4,\beta} =1$, is of interest, in order to work out the proof.
 Then eq.\ref{HSr} entails that: 
\begin{equation} 
\label{t4}
(H'-\epsilon_2)(a_1\Phi_{1,\beta}+a_2\Phi_{2,\beta})=0\hspace{1ex}
\Leftrightarrow \hspace{1ex}
\left((H'-\epsilon_2)(a_1b^{+}(k_1,k_2)|0\rangle+
a_2b^{+}(k_3,k_4)|0\rangle)\right)\otimes\Phi_{12}=0
\hspace{1ex}.
\end{equation} 
Because of $\Phi_{12}\ne 0$, eq.\ref{t4} implies that each eigenvalue 
$\epsilon_2$ of $H'$ in 
$S_2$ is also an eigenvalue of $H'$ in the subspace of Slater 
determinants
 made up of a single pair, the dimension $d_s$ of which is equal to 
$N$ or $N/2$ according to $\zeta=0$ or $\zeta=\pm 1$. It ensues that 
the dimension $d_2$ of $S_2$ is such that $d_2\le d_s$, which 
is in contradiction with the inequality $d_s<<d_2$ resulting 
from the fact that  $S_2$ 
is spanned by Slater determinants made up of $n$ pairs with $n>>1$.
 In addition eq.\ref{t3} yields  
$\langle\Phi_p|H_D|\Phi_p\rangle=\epsilon_2$ for every $\Phi_p$ 
 making up the linear expansion of $\Psi_2$ in 
eq.\ref{t1}. {\bf Q.E.D.}\par
	As every off-diagonal term $\langle\Phi_p|H'|
\Phi_q\rangle$ vanishes for the $\Phi_p,\Phi_q$ states 
 coming up in the linear expansion of $\Psi_2$ in 
eq.\ref{t1}, the off-diagonal and real space long-range 
order parameters 
in eqs.\ref{OD},\ref{RS} reduce both 
for $\psi_2$ to a two-particle distribution function:
\begin{equation}
\label{D}
\begin{array}{c}
f_{odlro}(|\tau|)=\sum_{k,k',\sigma}\left(\cos((k+k').\tau) 
-\cos((k+k').\tau+(k-k').\rho)\right)\langle\psi_2|
c^+_{k,+}c_{k,+}
c^+_{k',\sigma}c_{k',\sigma}|\psi_2\rangle
\hspace{1ex},\\
f_{rslro}(|\tau|)=\sum_{k,k',\sigma}\cos((k-k').
\tau)\langle 
\psi_2|c^+_{k,+}c_{k,+}
c^+_{k',\sigma}c_{k',\sigma}|\psi_2\rangle\hspace{1ex}.
\end{array} 
\end{equation} 
As a consequence of Riemann-Lebesgue's 
Theorem due to   
the oscillating character of $\cos((k\pm k').\tau)$, 
$f_{rslro}(|\tau|)$ and $f_{odlro}(|\tau|)$, calculated 
with $\rho$ kept fixed,     
decay towards zero for $|\tau|\rightarrow\infty$ so that 
$\psi_2$ has neither off-diagonal 
nor real space long-range order. Furthermore they may  
behave like power laws for large $|\tau|$ in similarity with 
previous results worked out in one dimension\cite{shi}. 
	       \section{Conclusion}
  	The general hamiltonian $H$ of
eq.\ref{h0}  has been shown to have two types 
of  eigenstates and eigenvalues $\psi_1,\epsilon_1$ and 
$\psi_2,\epsilon_2$ 
in the space of Slater determinants.
 The  $\psi_1$'s are characterised by a non vanishing 
projection in the space 
$S_{K,\zeta}$, denoted $\psi_{K,\zeta}$. This latter 
is responsible for off-diagonal long-range order and fulfils $(H_D+
H_{K,\zeta}-\epsilon_1)\psi_{K,\zeta}=0$ whereas  the  $\psi_2$'s 
obey  $(H_D-\epsilon_2)\psi_2=0$ and do not have 
off-diagonal long-range order. 
These results are valid  for arbitrary crystal
dimension, electron concentration and two-electron coupling 
provided it conserves $K$ and $\zeta$ in a scattering process.\par
	To realize that off-diagonal long-range order and real
space long-range order have different properties, it is 
illuminating to discuss 
the simple case of two electrons coupled by a one-dimensional 
Hubbard hamiltonian \cite{h1}. This system sustains a single
 band of 
bound eigenstates $\psi_{K,\zeta=0}$. As $\psi_{K,\zeta=0}$ is 
$\psi_1$-like, its off-diagonal long-range order parameter 
$f_{odlro}(|\tau|)$ oscillates 
like $\cos(K.\tau)$ while its 
real space long-range order parameter $f_{rslro}(|\tau|)$ 
decays like $e^{-\frac{|\tau|}{l}}$
where $l$ represents the size of the bound electron pair. 
 Each $\psi_{K,\zeta=0}$ is thus seen to have  
off-diagonal long-range order 
but no real space long-range order. Note also that 
Mermin-Wagner's Theorem 
\cite{aue} which rules out the possibility
 of real space long-range order in one and two dimensions is 
not relevant to Theorems 1 and 2 because  this  statement 
 is based actually  on a thermal average which 
fails to say anything upon the correlation properties
 of the many-body eigenstates. Thus if an eigenstate happens 
to have long-range order of any kind, 
 Mermin-Wagner's Theorem merely says that its statistical 
weight in the thermal average is too weak to give rise to  
 long-range order at finite temperature in the whole electron 
system. 
 Anyhow as the  eigenstates of all interacting 
electron systems, thus including  
 metals with long-range magnetic order but finite resistivity,  
have been shown to be either $\psi_1$- or $\psi_2$-like, 
Theorem 1 ensures 
that off-diagonal long-range order is 
not a sufficient criterion for 
superconductivity.\par
	    As far as  $\psi_1$-like solutions are concerned, 
the results of this work enable one to diagonalise 
$H$ on a cluster of size considerably 
larger than currently reached, because the dimension of 
$S_{K,\zeta}$ 
is much smaller than that of $S_\phi$. At last they 
provide useful constraints on
the variational states currently used in the many-body
problem such as those quoted in section 1. Indeed any 
variational state within the frame of reference of this work 
which is recognised to be   
neither $\psi_1$- nor $\psi_2$-like is unphysical even 
though it may fortuitously approximate the 
groundstate energy. In particular as the groundstate
energies of the one- and two-dimensional Hubbard hamiltonians are
smaller \cite{shib,dag} than the lowest eigenvalue of $H_D$,
  the respective groundstates are inferred to be  $\psi_1$-like.
 Since the spectrum of eigenvalues of $H$ has been shown 
to include all eigenvalues 
of every $H_{K,\zeta}$, and the BCS hamiltonian is equal 
to  $H_{K=0,\zeta=0}$ in the particular case where the 
Hubbard hamiltonian is equal to $H$,
  the Hubbard  hamiltonian, currently used to study the normal state, 
  turns 
out  to account for the properties of the superconducting state too. 
Finally this work provides 
a unified picture for the electron interaction in solids, valid for normal, 
magnetic and superconducting metals as well.\par 
	We are greatly indebted 
to J. Hlinka, P. Laurent, P. L\'ederer, M. Lewenstein, H. Moudden, 
A. Ole\'s and S. Petit 
 for helpful 
comments. 
 One of us (J.S.) dedicates this work to the memory 
of his parents Jochweta and Chaim and  his niece Denise L\'evy 
and thanks his 
wife Rachel and children J\'er\'emie and Judith for providing 
encouragement.
 \begin{center}
	{\bf Appendix}\\
 \end{center}
In the four electron case $(n=2)$, a  Slater determinant  
reads 
$\phi=c^+_{k_1}c^+_{k_2}c^+_{k_3}c^+_{k_4}|0\rangle$ 
where $k_1,k_2,k_3,k_4$ are four
vectors of the Brillouin zone and the spin indices $\sigma,\zeta$ are
dropped for simplicity in this example. An application of
eqs.\ref{30},\ref{exp} yields $m=3$ and
$\Phi_{\alpha=1}=b^+(k_1,k_2)|0\rangle\otimes b^+(k_3,k_4)|0\rangle,
\Phi_{\alpha=2}=b^+(k_1,k_3)|0\rangle\otimes b^+(k_4,k_2)|0\rangle,
\Phi_{\alpha=3}=b^+(k_1,k_4) b^+(k_2,k_3)|0\rangle$ and
$\Phi=(\Phi_1+\Phi_2+\Phi_3)$, if it is assumed 
that $k_1+k_2\ne
k_3+k_4, k_1+k_3\ne k_4+k_2, k_1+k_4=k_2+k_3$.
 The pair configurations $\alpha=1,2,3$  are characterised by 
the pair numbers 
$n_{k_1+k_2,\alpha=1}=n_{k_3+k_4,\alpha=1}=n_{k_1+k_3,\alpha=2}
=n_{k_4+k_2,\alpha=2}=1,n_{k_1+k_4,\alpha=3}=2$.  
Notice that $\Phi_1,\Phi_2,\Phi_3$ are three
linearly independent vectors of  $S_{\otimes\phi}$, all associated 
with the
same  vector $\Phi$ of $S_\Phi$ or equivalently $\phi$ of
$S_\phi$.

\end{document}